\documentstyle[prl,aps,floats]{revtex}
\input epsf

\begin{document}

\draft
\newcommand{\vertsp}{\vphantom{\displaystyle{\dot a \over a}}}
\newcommand{\Spin}[4]{\, {}_{#2}^{\vphantom{#4}} {#1}_{#3}^{#4}}

\title{Cosmic microwave background anisotropy from wiggly strings}


\author{Levon Pogosian and
Tanmay Vachaspati}
\address{
Physics Department, 
Case Western Reserve University, 
Cleveland OH 44106-7079.
}


\wideabs{

\maketitle


\begin{abstract}
\widetext
{\it April 6, 2006. Note: some normalization errors were discovered in our computer code. In order to correctly interpret numerical results in this paper please see \cite{PWW06}.}
We investigate the effect of wiggly cosmic strings on the cosmic 
microwave background radiation anisotropy and matter power spectrum 
by modifying the string network model used by Albrecht {\it et al.}. 
We employ the wiggly equation of state for strings and the one-scale 
model for the cosmological evolution of certain network characteristics. 
For the same choice of simulation parameters we compare the results with 
and without including wiggliness in the model and find that wiggliness 
together with the accompanying low string velocities lead to a 
significant peak in the microwave background anisotropy and to an 
enhancement in the matter power spectrum. For the cosmologies we 
have investigated (standard CDM, and, CDM plus
cosmological constant), and within the limitations of our modeling 
of the string network, the anisotropy is in reasonable 
agreement with current observations but the COBE normalized amplitude 
of density perturbations is lower than what the data suggests. In
the case of a cosmological constant and CDM model, a bias factor of 
about 2 is required.
\end{abstract}


}

\narrowtext

\section{Introduction}
\label{introduction}

The origin of the large-scale structure of the Universe has been a
focus of active research for the last few decades and is still 
one of the major unresolved problems in cosmology.
Measurements of the anisotropy in the Cosmic Microwave Background Radiation
(CMBR) have inspired hope that different theories of structure formation 
can be tested observationally. Quantum fluctuations in inflationary
models and topological defects are two important sources of density
inhomogeneities that would have different imprints on the CMBR
and which one could hope to distinguish.

Cosmic strings are a special variety of topological defects
that have been studied in the context of seeding large-scale
structure. Rapid progress on both observational and
theoretical issues has resulted in more accurate estimates
of the CMBR anisotropy due to strings 
\cite{penseltur,aletal,aletal2,hind98,aveletal,copmagst}. The
existence of publicly available computer codes \cite{selzal}
that can determine the anisotropy for given matter energy-momentum
tensors has greatly facilitated the study of mechanisms that can
source the CMBR anisotropy.

The major hurdle in calculating the effect of strings on the CMBR 
is that there is no simple way to characterize the network of 
strings. Large scale computer simulations have provided insight
into some of the properties of the network \cite{shelal}
within some choice of background cosmologies. However, there
are still unknown characteristics that could be important for
predictions of the CMBR anisotropy and the large-scale power
spectrum. As a result, certain extrapolations
have to be made to characterize the string network in the regime
where details are not yet available. This ``network modeling''
is the crucial aspect
of analyzing the observable signatures of cosmic strings.


Over the last decade or so, a number of approaches have been taken
to determine the detailed imprint of cosmic strings on the CMBR. 
The first attempt in this direction worked directly with the network
of strings found in computer simulations \cite{nature}. 
Recently, this numerical analysis has been refined and extended 
\cite{All7}.
The drawback in any such numerical attempt is that the dynamic range
of the simulations cannot yet extend to cover a cosmic expansion factor 
of at least $10^4$. It appears that some network modeling is essential
as was first attempted in \cite{Per}. 
In recent work \cite{hind98} the 
CMBR anisotropy was calculated using the results of a lattice simulation 
of a string network in a flat spacetime background.
Another approach, first suggested in \cite{hind97} and further developed
in \cite{aletal,aletal2}, approximates the string network 
by a collection of randomly oriented straight string segments, moving with
random velocities. This model has the merit of being relatively simple
and amenable to modifications that seem to be indicated by direct
simulations. Hence, we have adopted this model and, for the first
time, included small-scale structure on the string segments. Furthermore, 
we have modeled the parameters of the segments (length, velocity, 
wiggliness) by using the ``one-scale model'' of cosmic string 
networks \cite{Kib,Ben}
 and some reasonable expectations of how the 
strings would behave in an inflating background.

The most important feature that we have
taken into account in modeling the string network is the presence of 
small-scale structure on the strings. The scale of the
wiggles is much smaller than the characteristic length of the string. In
fact, a distant observer will not be able to resolve the details of the
wiggly structure. Instead, s/he would see a smooth string, but
the string would possess somewhat different characteristics. 
For a string with no wiggles, the equation of state is
$\mu=T=\mu_0$, where $\mu$ is the effective
mass per unit length and $T$ is the tension of the string. A wiggly string,
however, has a different effective equation of state:
\cite{carter,avilen},
$$ 
\mu T = \mu_{0}^2, \ \  \mu > T \ .  
$$
The wiggly string is heavier and slower than a Nambu-Goto string. 
 
The perturbation in the metric produced by a wiggly string is similar 
to that of a smooth string. As was shown in \cite{tvav}, in both
cases, the space in the vicinity of a straight segment of a string
is conical with a deficit angle $\Delta = 8 \pi G \mu$. The deficit 
angle is larger for wiggly strings, since they have a larger effective
mass per unit length. However, a new important feature of the wiggly 
string metric is a non-zero Newtonian potential 
$\phi \propto (\mu-T)$. (So the wiggly string behaves as a superposition
of a massive rod and a string with a conical deficit.)
As a result, massive particles will experience
a new attractive force $F \propto (\mu - T)/v_s$, where
$v_s$ is the string velocity \cite{tvav}.
At the same time, the propagation of photons in a plane perpendicular
to the string is unaffected by the rod-like feature of the wiggly
string metric and is only affected by the conical deficit.
This fact - that the effect of wiggliness on massive particles is 
qualitatively different from the effect on radiation - seems especially 
relevant when calculating CMBR anisotropy and the power spectrum.
It may be hoped that this feature could help alleviate
the current difficulties in reconciling the COBE normalized matter 
power spectrum with the observational data in the cosmic string
model.

It should be pointed out that the model we have adopted cannot be
the final word since a number of factors have still not been taken
into account. In the network itself, the infinite strings will
produce loops which will then decay by emitting gravitational
radiation. The effect of the loops is only included insofar that they
can be treated as segments of strings. In other words, large loops
are included in our analysis but small loops have been missed.
Also, the decay of loops into gravitational radiation has been
missed. The back-reaction of gravitational radiation on the string
segments is completely neglected. We hope to rectify some of these
omissions by further modeling of the network in subsequent work.

In the following sections we use the model developed in
Ref. \cite{aletal} to calculate the CMBR anisotropy and the
matter power spectrum due to cosmic strings in a few cosmologies. 
The energy-momentum tensor
that we use is that of wiggly strings, as we describe in the
next section. In addition, the characteristics of the string
network (velocity and correlation length) are computed using the
one-scale model described in Sec. \ref{network}. There are some
further assumptions that we have to make about the string
network that we also describe in this section. Our results for
the CMBR anisotropy and the power spectrum in density inhomogeneities
for a few different cosmologies are given in Sec. \ref{results}.
Here we also discuss the model dependence of our results. 
Our conclusions are summarized in Sec. \ref{conclude}.

\section{Wiggly strings}
\label{wigglystrings}

The world history of a string can be represented by a two-dimensional
surface in spacetime,
\begin{eqnarray*}
x^\mu=x^\mu(\zeta^a),\; a=0,1,
\end{eqnarray*}
called the string worldsheet, where $\zeta^0$ and $\zeta^1$ are the
coordinates on the worldsheet.

We consider strings that live in an expanding universe described by
the metric,
\begin{eqnarray*}
ds ^2 = a ^2 (\tau) (d \tau ^2 - d {\bf x} ^2 ),
\end{eqnarray*}
where $\tau$ is the conformal time. The string equations of motion are 
invariant under reparametrization of the string worldsheet.
A convenient choice of the parametrization (gauge) is
\begin{eqnarray*}
\zeta^0=\tau \, ,\ {\bf x} ' \cdot \dot{\bf x}=0,
\end{eqnarray*}
where the prime denotes a 
derivative with respect to $\zeta^1$.
In this gauge the energy-momentum tensor of a string is
\begin{equation}
T^{\mu\nu}(y)= \frac{\mu}{\sqrt{-g}} \int d^2 \zeta
(\epsilon \dot{x}^\mu \dot{x}^\nu- \epsilon^{-1}
x^{'\mu} x^{'\nu}) \delta^{(4)} (y - x(\zeta)),
\label{emt}
\end{equation}
where $\sigma=\zeta^1$ and
\begin{equation}
\epsilon = \sqrt{ \frac{{\bf x}^{'2}}{1-\dot{\bf x}^2} }.
\label{epsilon}
\end{equation}

Following Carter \cite{carter} we can define the string tension 
$T$ and the  string mass-energy per unit length $U$ by 
\begin{equation}
\sqrt{-g} T^{\mu\nu}(y)=
\int d^2 \zeta \sqrt{- \gamma} ( U u^\mu u^\nu
- T v^\mu v^\nu)
\delta^{(4)} (y - x(\zeta)),
\label{tension}
\end{equation}
where $u^\mu$ and $v^\mu$ are such that 
\begin{eqnarray}
u_\mu u^\mu=-v_\mu v^\mu =1,\;  u_\mu v^\mu =0, \,\label{conda}\\
(u^\mu v^\rho - v^\mu u^\rho) (u_\rho v_\nu - v_\rho u_\nu) =
\eta ^\mu _\nu, \,\label{condb}\\
\eta ^{\mu\nu} = \gamma ^{ab} x^{\mu}_{,a} x^{\nu}_{,b},
\,\label{condc}
\end{eqnarray}
and $\gamma^{ab}$ is the worldsheet metric.
One can check that 
\begin{equation}
u^\mu=\frac{\sqrt{\epsilon}\dot{x}^\mu}{(-\gamma)^{1/4}},\;
v^\mu=\frac{x^{'\mu}}{\sqrt{\epsilon} (-\gamma)^{1/4}}
\label{defuv}
\end{equation}
satisfy the conditions (\ref{conda})-(\ref{condc}).
Substituting (\ref{defuv}) into (\ref{emt}) and comparing with
(\ref{tension}) one obtains
\begin{equation}
U=T= \mu,
\label{state0}
\end{equation}
which is the equation of state for an ordinary (``intrinsically
isotropic'' \cite{carter}) Nambu-Goto string \cite{vilen85}.

Lattice simulations of string formation and evolution suggest
that strings are not straight but have wiggles \cite{shelal}. 
A distant observer, however, 
would not be able to resolve the small-scale structure. Insead, he 
would see a 
smooth string with the effective mass per unit length $\tilde{U}$ and the 
tension $\tilde{T}$. The equation of state for a wiggly string averaged
over the small-scale structure has been shown \cite{carter}
\cite{avilen} to be
\begin{equation}
\tilde{U} \tilde{T} = \mu^2,
\label{state1}
\end{equation}
or
\begin{equation}
\tilde{U}=\alpha\mu,\;  \tilde{T} = \mu / \alpha,
\label{state2}
\end{equation}
where $\alpha=\alpha(\sigma,\tau)$ is, in general, some function
of time and the coordinate along the length of the smoothed string. 
The energy-momentum tensor of a wiggly
string viewed by an observer who cannot resolve the wiggly structure
is then obtained by substituting (\ref{state2}) into (\ref{tension}):
\begin{eqnarray*}
\tilde{T}^{\mu\nu}(y)=
\frac{1}{\sqrt{-g} }\int \frac{d^2 \zeta}{\sqrt{- \gamma}}
( \tilde{U} u^\mu u^\nu
- \tilde{T} v^\mu v^\nu)
\delta^{(4)} (y - x(\zeta)) \\
=\frac{\mu}{\sqrt{-g}} \int d^2 \zeta  
\left[\epsilon \alpha \dot{x}^\mu \dot{x}^\nu- \frac{
x^{'\mu} x^{'\nu}}{\epsilon \alpha}\right]
\delta^{(4)} ( y -  x(\sigma ,\tau)).
\end{eqnarray*}

Next, we use this expression to calculated the stress-energy of a string
network.

\section{The string network}
\label{network}

\subsection{Parameters of the network}
\label{networkparameters}

The string network at any time can be characterized by a single length 
scale, the correlation length $L$, defined by
\begin{equation}
\rho=\frac{\mu}{L^2},
\label{corlen}
\end{equation} 
where $\rho$ is the energy density in the string network.
It will be convenient to work with the comoving correlation 
length ${\it l}=L/a$.

The expansion stretches the strings, thus increasing the energy density.
At the same time, long strings reconnnect and chop off loops which
later decay. The evolution of ${\it l}$ including only these two competing
processes is \cite{Kib,Ben,marshe}
\begin{equation}
\frac{d{\it l}}{d\tau}=\frac{\dot{a}}{a}{\it l}v^2+\frac{1}{2}\tilde{c}v \\
\label{evol1}
\end{equation}
\begin{equation}
\frac{dv}{d\tau}=(1-v^2) ( \frac{\tilde{k}}{{\it l}}-2\frac{\dot{a}}{a}v ),
\label{evol2}
\end{equation}
where $v$ is the rms string velocity, 
$\tilde{c}$ is the loop chopping efficiency and $\tilde{k}$ is
the effective curvature of the strings. 
The values of $\tilde{c}$ and $\tilde{k}$ in radiation and matter eras
are suggested in \cite{marshe}. We use the same
scheme as in 
Ref. \cite{aletal2} to interpolate between these values through the
radiation-matter transition:
\begin{eqnarray*}
\tilde{c}(\tau)=\frac{c_r+g a c_m}{1+g a} \\
\tilde{k}(\tau)=\frac{k_r+g a k_m}{1+g a},
\end{eqnarray*}
where we take $c_r=0.23$, $c_m=0.18$, $k_r=0.17$, $k_m=0.49$, $g=300$
and $a(\tau)$ is normalized so that $a=1$ today\footnote{
This choice of the value of $g$ along with our normalization for $a$
leads to the same time-dependence of $v$ and $l$ as in \cite{aletal2}. 
Normalizing $a$ so that $a=1$ at equality would require a much smaller
value, e.g. $g \sim 0.1$, as reported in \cite{aletal2}}.

\subsection{Model of the network}
\label{networkmodel}

Here we are closely following the model described in 
\cite{aletal,aletal2}. 

\begin{figure}
\centerline{\epsfxsize = 0.8\hsize \epsfbox{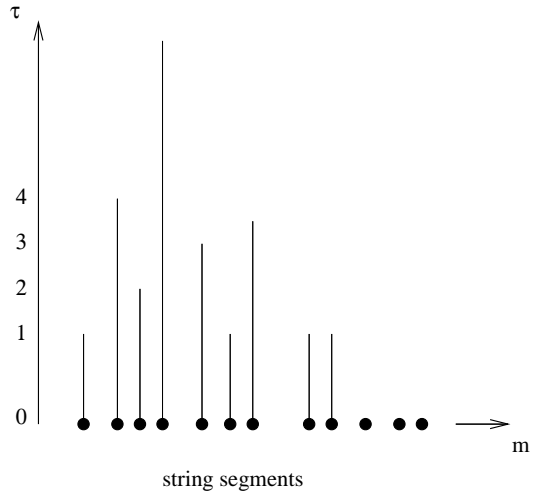}}
\vskip 0.5 truecm
\caption{\label{net1}
A schematic picture of the string network model. All string
segments (depicted by filled circles)
are born at an early epoch and then decay at various 
later times. The segments are labeled by the index $m$ and the
decay times are shown as numbers along the $\tau$ axis. 
In certain cosmologies, it is possible that some string segments 
never decay. 
}
\end{figure}

The basic picture is that the string network is represented
by a collection of uncorrelated, straight string segments moving
with random, uncorrelated velocities.
All the segments are assumed to be produced at some early epoch.
At every subsequent epoch, a certain fraction of the 
number of segments decay
in a way that maintains network scaling. 
This picture of the network is depicted in Fig. \ref{net1}.

The comoving length, $l$, of each segment at any time
is taken to be equal to the correlation length of the network 
defined below eq. (\ref{corlen}). The positions of the segments
are drawn from a uniform distribution in space and
their orientations are chosen from a uniform distribution 
on a two sphere. 
The segment speeds are fixed to be given by the solution
of eq. (\ref{evol2}) while
the direction of the velocity is taken to be uniformly
distributed in the plane perpendicular to the string 
orientation\footnote{We have also performed a 
few simulations where we drew the velocities from a Gaussian 
distribution as in Ref. \cite{aletal}, but these did not lead 
to significantly different results.}.
In principle, this constraint on the velocity does not remain
valid when the strings are wiggly since the wiggles can impart
a longitudinal velocity to the segments. However, 
as explained below, the longitudinal
velocities are expected to be much smaller than the transverse
velocities and hence will be neglected.

The decay of segments of the string network is accomplished by 
``turning off'' the energy-momentum of a fraction of the existing segments
at every epoch. 
Each segment is assigned a certain decay time, $\tau_m$, where the index 
$m$ labels the individual segments. So the Fourier transform of the 
total stress-energy of the network is the sum over the stress-energies 
of all segements:
\begin{equation} 
\Theta_{\mu\nu} ( \vec{k} ,\tau ) =
\sum_{m=1}^{N_0} \Theta^m_{\mu\nu}  ( \vec{k} ,\tau ) 
T^{\rm off} \left(\tau,\tau_m\right)\,,
\label{thetamunu}
\end{equation} 
where $N_0$ is the initial number of segments, and
$T^{\rm off} (\tau, \tau_m)$ is a smooth function that turns off 
the $m^{th}$ string segment by time $\tau_m$~\footnote{
In addition, Albrecht {\it et al.} introduce a function $T^{on}$ which
turns on the segments at a very early time. This function is not a
feature of the model but only introduced to speed up the code. We
have not included $T^{on}$ in our simulations.}.
The functional form is taken to be \cite{aletal}
\begin{equation}
T^{\rm off}(\tau,\tau_m) = \left\{ \begin{array} {ll}
         1  & \ldots \tau < L\tau_m \\
         \frac{1}{2}+\frac{1}{4}(x^3-3x) & \ldots L \tau_m< \tau < \tau_m \\
         0  & \ldots \tau > \tau_m
         \end{array} \right.,
\label{toff}        
\end{equation}
where 
\begin{equation}
x=2 \frac{\ln (L \tau_m/\tau)} {\ln (L)} -1\,
\end{equation}
and $L<1$ is a parameter that controls how fast the segments decay.

The total energy of the string network in a volume $V$ at any time is
\begin{equation}
N \mu L =
V\rho=\frac{\mu V}{L^2}
\label{vrho}
\end{equation}
where $N=N(\tau )$ is the total number of string segments at that time,
$V=V_0 a^3$, $a=1$ at the present epoch and $V_0$ is a constant 
simulation volume. From (\ref{vrho}) it follows that
\begin{equation}
N=\frac{V}{L^3}=\frac{V_0}{{\it l}^3}.
\end{equation}

The comoving length ${\it l}$ is approximately proportional to the conformal 
time $\tau$ and implies that the number of strings $N(\tau)$
within the simulation volume $V_0$ falls as $\tau^{-3}$. To calculate the CMB
anisotropy we need to evolve the string network over at least four orders
of magnitude in cosmic expansion. Hence we would have to start with $N \gtrsim
10^{12}$ string segments in order to have one segement left at 
the present time. This is the main problem in directly dealing with the
expression for the energy momentum tensor in eq. (\ref{thetamunu}).

\begin{figure}
\centerline{\epsfxsize = 0.7\hsize \epsfbox{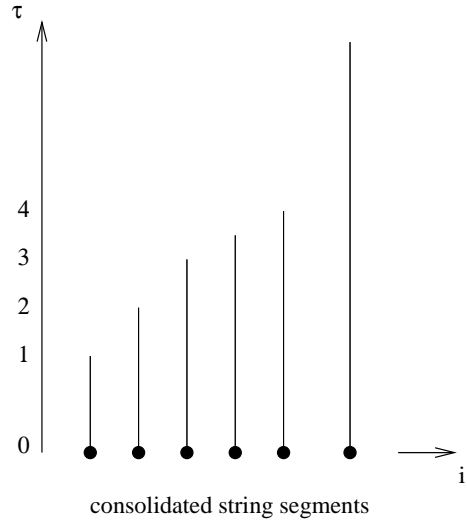}}
\vskip 0.5 truecm
\caption{
\label{net2}
The modified model of the string network. All strings that
decay at the same (discretized) time in Fig. 
\protect \ref{net1}
are consolidated into one
string segment and assigned a weight that is the square root
of the number of segments that the consolidated segment
represents. This works for all segments that decay by the
end of the simulation but will miss those segments that do not
decay. In the present scheme, the segments that will not have
decayed by the end of the simulation are consolidated into
one string. The contribution of this surviving segment is the
remainder term in the expression for the energy momentum
tensor in eq. 
(\protect \ref{modifiedtheta}).
}
\end{figure}

A way around this difficulty was suggested in Ref. \cite{aletal}. The 
suggestion is to consolidate all string segments that decay at the
same epoch. Since the number of segments that decay at the
(discretized) conformal time $\tau_i$ is
\begin{equation}
N_d (\tau_i ) = V [ n(\tau_{i-1}) - n(\tau_i) ]
\label{decayingstrings}
\end{equation}
where $n(\tau )$ is the number density of strings at time $\tau$,
the ``consolidated string'' decaying at $\tau_i$ is taken to have 
weight: $\sqrt{N_d (\tau_i )}$. (The assumption is that string
segments that decay at the same time
act randomly, leading to the square root in the weight.) 
The number of consolidated strings is of the order of a few hundred 
and can be dealt with computationally\footnote{In addition to consolidating
the string segements decaying at a given time into a single 
``consolidated string''
we have also tried consolidating them into two, three and four strings.
This did not lead to any difference in the final results.}.
This modified picture of the string network is shown in Fig. \ref{net2}.

Now we can choose $\tau_n$ to be equally spaced on a logarithmic scale 
between $\tau_{min}$ and $\tau_{max}$ and write the energy-momentum tensor 
as
\begin{equation}
\Theta_{\mu\nu} ( \vec{k} ,\tau )=
\sum_{i=1}^K [ N_d(\tau_i ) ]^{1/2} 
 \Theta^i_{\mu\nu}  ( \vec{k} ,\tau )
  T^{\rm off} \left(\tau,\tau_i \right)  + {\cal T}_{\mu \nu}\,.
\label{modifiedtheta}
\end{equation}
where $K$ is the number of consolidated segments and ${\cal T}_{\mu\nu}$ is
a remainder that we have included and that we now explain.

The sum in eq. (\ref{modifiedtheta}) misses any string segments
that have not decayed by $\tau_{max}$. The remainder ${\cal T}_{\mu\nu}$
is supposed to represent the contribution of these segments of 
strings. As in the case of the decaying strings, we will
consolidate these surviving segment into one segment with the
weight $\sqrt{V n(\tau_{max} )}$. Therefore
\begin{equation}
{\cal T}_{\mu\nu} = 
\sqrt{V n(\tau_{max} )} ~ \Theta^{(1)}_{\mu\nu} ( \vec{k} ,\tau )
\label{remainder}
\end{equation}
where, the superscript on $\Theta$ means that it is the energy-momentum
tensor for one string segment.

For certain cosmologies - those without a cosmological constant -
the remainder term can be made arbitrarily small by choosing a large 
enough $\tau_{max}$. This is because $n(\tau ) \propto l(\tau )^{-3}$,
where $l(\tau ) \sim \tau$ is the comoving correlation length,
and hence $n (\tau) \rightarrow 0$ as $\tau \rightarrow \infty$. 
However, if the cosmological constant
is non-zero, the universe enters an inflationary epoch and the
range of the conformal time is finite, $\tau \in [0, \tau_\infty ]$,
and given by:
$$
\tau_\infty = \int_0^{\tau_\infty} d\tau = 
\int_0^\infty {{dt} \over {a(t)}}  \ .
$$

In this model, the 
number density of string segments is assumed to only depend on the
correlation length as $n(\tau ) \propto l(\tau) ^{-3}$. However,
the decay of string segments as described by the
function $T^{off}$ prevents this relationship from being exact.
Instead we have:
$$
n(\tau ) = {{C(\tau )} \over {l(\tau )^3}} \ .
$$
The function $C(\tau )$ is determined by requiring that the
total number of strings at any time is still given by $V/l(\tau )^3$.
Therefore:
\begin{eqnarray*}
{1 \over {l(\tau )^3}} = 
\sum_{i=1}^K  [n(\tau_{i-1}) - n (\tau_i) ] 
T^{\rm off} (\tau,\tau_i) +  n(\tau_{max}) ,
\end{eqnarray*}
and $l(\tau )$ is determined from the one-scale model (eq. (\ref{evol1})).
(Note that $C(\tau )$ is contained in $n(\tau )$.)
It is reassuring to see that the function $C(\tau )$ that we obtain
from our code is nearly constant and of order unity throughout the 
simulation.

The Fourier transform of the energy-momentum tensor
of an individual string segment is
\begin{eqnarray*}
\Theta_{\mu\nu}(\vec{k},\tau)&=&
\int d^3 x e^{i\vec{k}\cdot\vec{x}} \Theta_{\mu\nu} \\
&=&\mu \int_{-{\it l}/2}^{{\it l}/2} d \sigma e^{i\vec{k}\cdot\vec{X}}
(\epsilon\alpha \dot{X}^\mu \dot{X}^\nu - \frac{1}{\epsilon\alpha}
X^{\prime\mu} X^{\prime\nu}).
\end{eqnarray*}
\begin{equation}
\label{emtft}
\end{equation}
where,
$X^\mu (\sigma,\tau)$ are the coordinates of the segment: 
\begin{equation}
X^0 = \tau ,\ \ 
\vec{X} = \vec{x}_0 + \sigma \hat{X}^\prime +v\tau \hat{\dot{X}},
\label{coord}
\end{equation}
where $\vec{x}_0$ is the random location of the center of mass, 
$\hat{X}^\prime$ and $\hat{\dot{X}}$ are randomly oriented unit
vectors satisfying $\hat{X}^\prime \cdot \hat{\dot{X}} =0$ and
$v$ is the velocity of the string. 

The random location vectors, $\vec{x}_0$, appear only in the dot 
product $\vec{x}_0 \cdot\vec{k}$. Instead of generating $\vec{x}_0$ 
at random, Albrecht {\it et al.} 
generate $\vec{x}_0 \cdot \vec{k}$ for each segment 
as a random number from $[0,2\pi]$. We have also followed this
scheme\footnote{In ongoing work, one improvement of the model
that we are investigating is a scheme in which we choose 
$k_{min} \vec{x}_0 \cdot \hat{k}$ at random in the interval 
$[0,2\pi ]$ and then multiply by $k/k_{min}$ 
where $k_{min}$ is the smallest wave-vector considered in the 
simulation.}. 

As mentioned earlier, we will be ignoring the longitudinal velocities
of the segments. This is the constraint 
$\hat{X}^\prime \cdot \hat{\dot{X}} =0$ that we have imposed. The
justification is that there are on the order of $10^4$ left- and
right- moving wiggles on a segment of string, each traveling with a 
velocity of about the speed of light. The average momentum of this
``gas'' of wiggles will be zero but there can be fluctuations which
will lead to a velocity $\sim 1/\sqrt{10^4}$. Such longitudinal
velocities are insignificant compared to the transverse velocity
expected to be $\sim 0.1$.

The differential equations which describe the metric perurbations 
produced by $\Theta_{\mu\nu}(\vec{k},\tau)$ do not depened on the direction 
of $\vec{k}$. Therefore, without any loss of generality we will assume 
$\vec{k}=\hat{k}_3 k$ in eq. (\ref{emtft}).

Substituting (\ref{coord}) into (\ref{emtft}), integrating over
$\sigma$ and taking the real part gives
\begin{eqnarray}
\Theta_{00} &=& \frac{\mu \alpha} {\sqrt{1-v^2}}
\frac{\sin(k\hat{X}_3^\prime{\it l}/2)}{k\hat{X}_3^\prime/2}
\cos(\vec{k} \cdot \vec{x}_0 + k\hat{\dot{X}}_3 v\tau) \,, \\
\Theta_{ij} &=& \left[ v^2 \hat{\dot{X}}_i \hat{\dot{X}}_j -
\frac{(1-v^2)}{\alpha^2} \hat{X}^\prime_i \hat{X}^\prime_j \right] \;
\Theta_{00} .
\label{emtcom}
\end{eqnarray}
$\Theta_{0i}$ can be found from the covariant conservation equations 
$\nabla^\mu \Theta_{\mu\nu}=0$. Here we have assumed that the
entire string energy is in long strings. The conservation is violated
if one includes the energy in loops and the gravitational radiation.
A detailed discussion of this issue can be found in \cite{hind98}.

\subsection{Evolution of parameters}
\label{parameterevolution}

\begin{figure}[tbp]
\centerline{\epsfxsize = 0.8\hsize \epsfbox{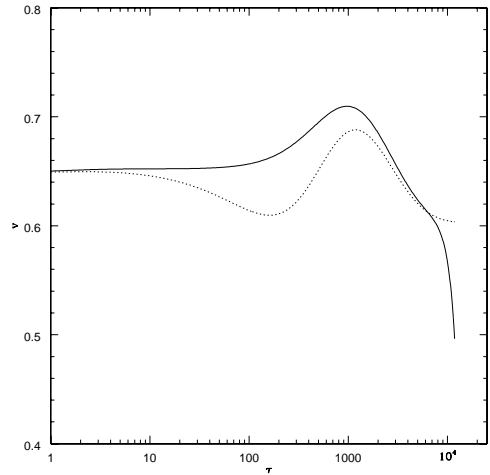}}
\vskip 0.5 truecm
\caption{ The velocity of strings $v$ as a function of the conformal time
for $\Omega_{baryons}=.05$, $\Omega_{CDM}=0.95$,
$\Omega_{\Lambda}=0$ (dotted line) and $\Omega_{baryons}=.05$, 
$\Omega_{CDM}=0.25$,
$\Omega_{\Lambda}=0.7$ (solid line).
} 
\label{fig-v}
\end{figure}

\begin{figure}[tbp]
\centerline{\epsfxsize = 0.85\hsize \epsfbox{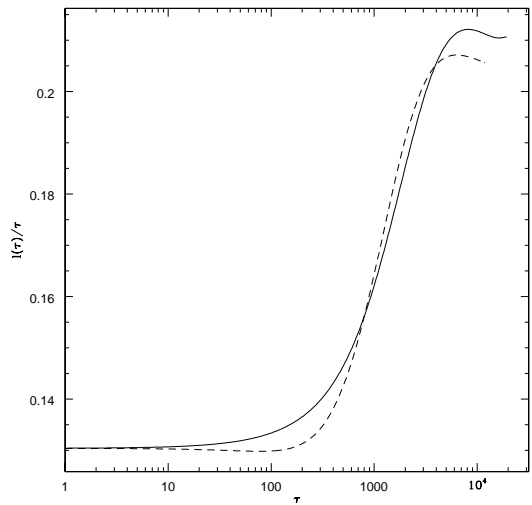}}
\vskip 0.5 truecm
\caption{The length of string segments, $l(\tau)$, divided by $\tau$ 
as a function of $\tau$ for $\Omega_{baryons}=.05$, $\Omega_{CDM}=0.95$,
$\Omega_{\Lambda}=0$ (dotted line) and $\Omega_{baryons}=.05$, 
$\Omega_{CDM}=0.25$, $\Omega_{\Lambda}=0.7$ (solid line).
} 
\label{fig-l}
\end{figure}

The evolution of the velocity, $v$, with time is found by
solving eqns. (\ref{evol1}) and (\ref{evol2}) with a range of possible intitial 
conditions. In Fig. \ref{fig-v} we show the 
evolution of the string velocity for the initial conditions:
$l(\tau_{min})=0.13\tau_{min}$, $v_0(\tau_{min})=0.65$ 
for two different cosmological models.
Similarly we show the behaviour of $l(\tau )/\tau$ as a
function of $\tau$ in Fig.\ref{fig-l}.

\begin{figure}[tbp]
\centerline{\epsfxsize = 0.85\hsize \epsfbox{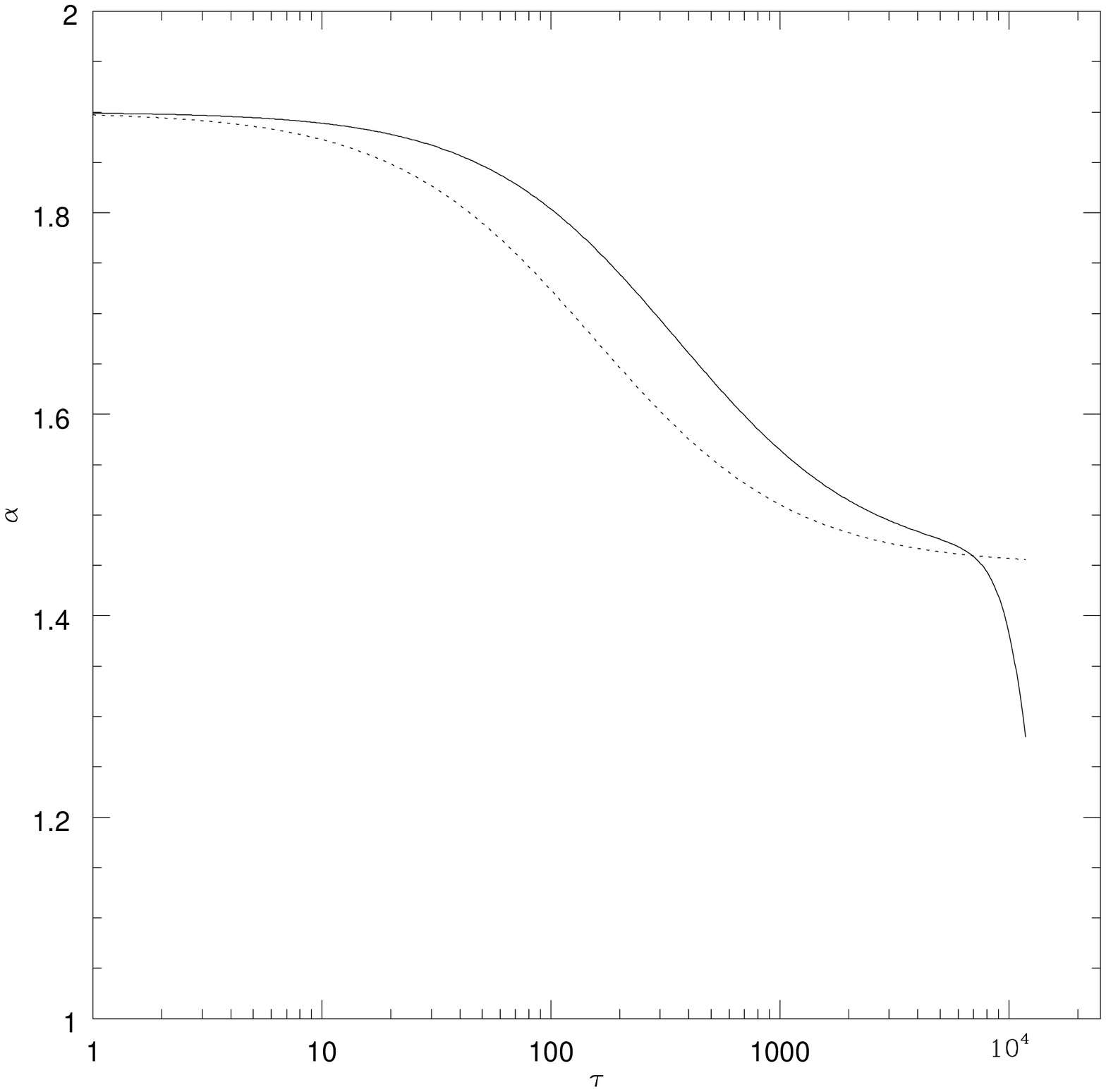}}
\vskip 0.5 truecm
\caption{ The wiggliness $\alpha$ as a function of the conformal time
for $\Omega_{baryons}=.05$, $\Omega_{CDM}=0.95$,
$\Omega_{\Lambda}=0$ (dotted line) and $\Omega_{baryons}=.05$, 
$\Omega_{CDM}=0.25$,
$\Omega_{\Lambda}=0.7$ (solid line).
}
\label{fig-alf}
\end{figure}

The values of the ``wiggliness'' parameter $\alpha$ have been 
estimated \cite{shelal} in the radiation and matter eras to be 
$\alpha_r \simeq 1.9$ and $\alpha_m \simeq 1.5$ 
respectively. In the case of a non-zero cosmological constant,
the exponential expansion of the Universe will stretch and smooth out 
the wiggles, so that 
$\alpha \rightarrow 1$ in the $\Lambda-$dominated epoch. A function that 
fits the expected evolution of $\alpha$ is (Fig. \ref{fig-alf})
\begin{equation}
\alpha(\tau)=1+\frac{(\alpha_r-1)a}{\tau \dot{a}} \ .
\end{equation}
We shall assume this behaviour of $\alpha$ in the next
section.

Next, we use the expression for the string stress-energy to calculate the
radiation and matter perturbation spectra at present time.

\section{Results}
\label{results}

We calculate the CMB anisotropy using the line of sight integration
approach \cite{selzal} implemented in the publicly available code
CMBFAST.  Given $\Theta_{\mu\nu}$ for each $k$ and $\tau$, 
CMBFAST integrates the Einstein equations simultaneously
with the Boltzmann equations for all radiation and matter particles
present in the model. Two factors that have not been included in
our analysis are: (1) compensation of string inhomogeneities by
perturbations in the matter fluids, and, (2) small loops of string.
The first factor was considered in Ref. \cite{aletal} which we
are closely following. However, it was found there that the effects
of compensation do not significantly alter the predictions of
anisotropy and power spectrum. It is possible that the second factor 
is important. At the moment, we do not have a good model for including
loops in our analysis as there is not much information available in
the literature on which to build such a model. The inclusion of loops
in the model is a problem that we postpone for the future, though
the slow decay of string segments described below may mimic the
presence of loops to some extent.

\begin{figure}[tbp]
\centerline{\epsfxsize = 0.8\hsize \epsfbox{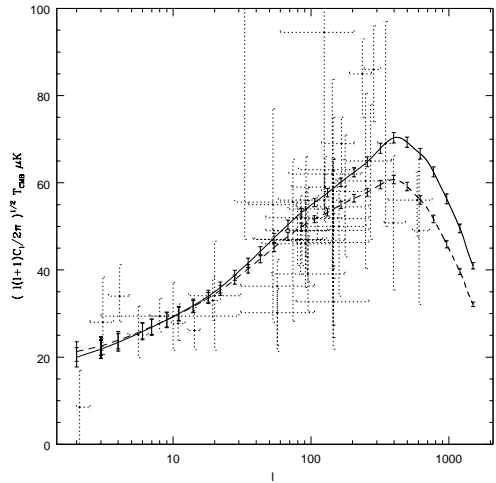}}
\vskip 0.5 truecm
\caption{
The total angular power spectrum with (solid line) and without (dashed line) 
including the wiggliness when $\Omega_{baryons}=.05$, $\Omega_{CDM}=0.95$, 
$\Omega_{\Lambda}=0$ and $v_0 = 0.65$.
The compiled observational data is plotted for comparison.
}
\label{fig-c00}
\end{figure}

\begin{figure}[tbp]
\centerline{\epsfxsize = 0.8\hsize \epsfbox{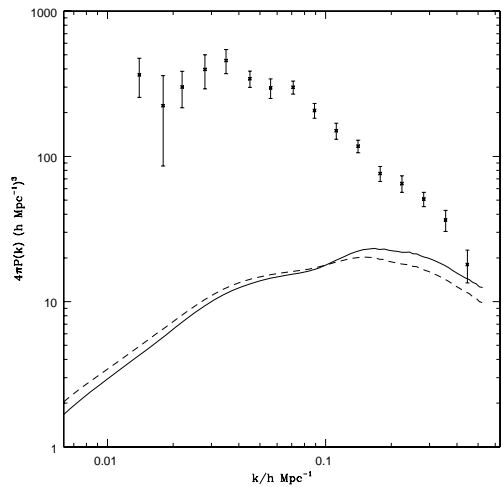}}
\vskip 0.5 truecm
\caption{
The matter power spectrum for
the same choice of parameters as in Fig. \protect \ref{fig-c00}.
The data extracted from surveys of galaxies and clusters of galaxies
is plotted for comparison.}
\label{fig-p00}
\end{figure}

The CMBR anisotropy as seen by an observer can be described by
$\Delta({\bf x}, \hat{n} ,\tau_o) \equiv |T({\bf x}, 
\hat{n} ,\tau_o)- \bar{T}|/\bar{T}$,
where ${\bf x}$ is the position of the observer, $\hat{n}$
is the line of sight direction, $\tau_o$ is the
conformal time today and $\bar{T}$ is the average temperature of 
CMBR. At ${\bf x}=0$ one can decompose $\Delta$
into spherical harmonics
\begin{equation}
\Delta(\hat{n})= \sum_{lm} a_{lm} Y_{lm} (\hat{n}).
\label{alm}
\end{equation}
The angular power spectrum $C_l$ is defined by
\begin{equation}
C_l \equiv \frac{1}{2l+1}\sum_{m=-l}^l \langle a_{lm}^{*} a_{lm} \rangle,
\label{defcl}
\end{equation}
where $\langle \rangle$ denotes an ensemble average.
A self-consistent treatment of the line of sight method and the 
complete set of the equations is given in \cite{whihu}.
The output of CMBFAST is
the COBE normalized angular power spectrum of temperature anisotropy
and the power spectra of each of the matter species that is present.

In its current version
CMBFAST calculates perturbations from scalar and tensor sources.
Since strings may generate a significant vector component, we had
to add the vector part to the code\footnote{Our code 
can be downloaded from the website:\\
http://theory4.phys.cwru.edu/$\sim$levon.}.

Our results for the anisotropy and power spectrum are averages over 
$M$ different realizations of a random network of strings. In the 
limit of large $M$, each quantity
that we calculate will have an associated probability distribution. 
For example, for each $l$ the distribution of $C_l$ will approach some 
probability function with mean $\bar{C_l}$ and standard deviation
$\sigma_l$. For the averaged result, $\bar{C_l}$, the standard
deviation is $\sigma_l /\sqrt{M}$.
Performing a large number of experiments on a computer allows us
to find $\bar{C_l}$ and $\sigma_l$ quite accurately. However,
an observer ({\it eg.} COBE) trying to determine $\bar{C_l}$ would only
be able to average over $(2l+1)$ causally disconnected patches on the
sky. Therefore, the accuracy in determining $\bar{C_l}$ observationally
is $\sigma_l/\sqrt{2l+1}$. We take this limitation (known as
``cosmic variance'') into account when plotting the $1\sigma-$error 
bars on our graphs.

Our results were found by averaging over $M=300$ string network 
realizations, however, 100 runs would be enough to reproduce most
of our results. The simulation with 100 runs took about 10 hours 
to run on a single processor IBM 590 workstation and about 5 hours
on 10 processors of the J90 cluster of the National Energy Research 
Scientific Computing Center.

We work only with a flat universe
with and without a cosmological constant. 
We consider two cases: ($\Omega_{baryons}=.05$, $\Omega_{CDM}=0.95$,
$\Omega_{\Lambda}=0$) and ($\Omega_{baryons}=.05$, $\Omega_{CDM}=0.25$,
$\Omega_{\Lambda}=0.7$). The first case is what until recently was known
as the standard CDM model. The second is motivated by the recent supernovae
data that suggest $\Omega_{\Lambda}=0.7$ and a flat universe.
The value of the Hubble constant in both cases was taken
to be $H_0 =50$ $km$ $sec^{-1}$ $Mpc^{-1}$.

In Fig. \ref{fig-c00} we show the effect that wiggliness has on
$C_l$'s for the case when $\Omega_{\Lambda}=0$ along
with the compiled experimental data \cite{data}. The matter power spectrum
for the same set of paramters is shown in Fig. \ref{fig-p00} together with
the data from surveys of galaxies and clusters of galaxies \cite{pedod}.
Adding wiggliness results in a higher peak at $l\approx 400$. However, it does
not improve the
shape or the magnitude of the matter power spectrum which appears to be in 
disagreement with data. The COBE normalization allows us to determine the 
string mass per unit length. We obtain that $G\mu_0=1.1\times 10^{-6}$ for
wiggly strings and $G\mu_0=1.5\times 10^{-6}$ for smooth strings.

\begin{figure}[tbp]
\centerline{\epsfxsize = 0.9\hsize \epsfbox{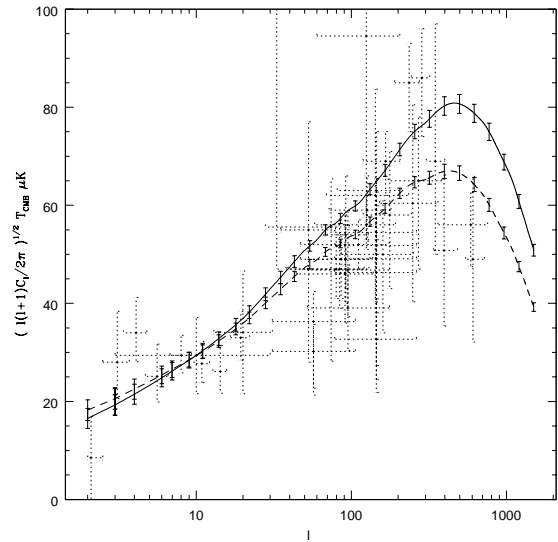}}
\vskip 0.5 truecm
\caption{
The total angular power spectrum for wiggly (solid line) and smooth 
(dashed line) strings when $\Omega_{baryons}=.05$, $\Omega_{CDM}=0.25$, 
$\Omega_{\Lambda}=0.7$ and $v_0 =0.65$.}
\label{fig-c70}
\end{figure}

\begin{figure}[tbp]
\centerline{\epsfxsize = 0.9\hsize \epsfbox{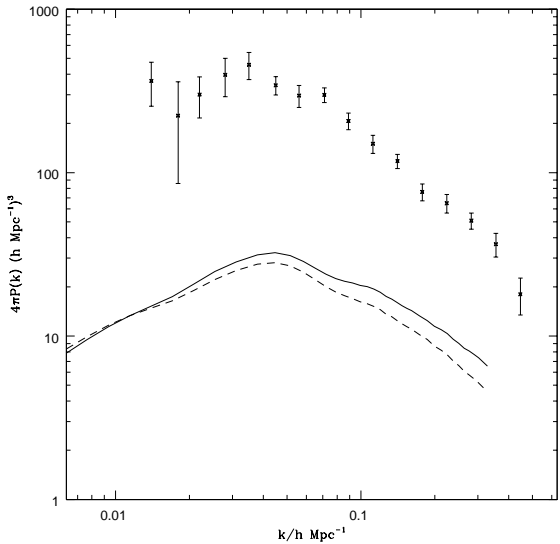}}
\vskip 0.5 truecm
\caption{
The matter power spectrum for
the same choice of parameters as in Fig. 
\protect \ref{fig-c70}.
}
\label{fig-p70}
\end{figure}

Allowing for a non-zero cosmological constant improves the agreement with
the data. In Fig. \ref{fig-c70} and Fig. \ref{fig-p70} we plot the results of
the simulation with $\Omega_\Lambda=0.7$. The peak in the CMBR anisotropy
is now higher compared to the $\Omega_\Lambda=0$ case.
The power spectrum in Fig. \ref{fig-p70} includes the factor of 
$\Omega_{matter}^{0.3}$ necessary for
correct comparison with data \cite{pedod}. Although the magnitude is 
significantly lower than the data, the shape of the matter power spectrum is
improved. The wiggliness results in a higher peak in $C_l$'s and a slight
increase in the magnitude of matter power spectrum at smaller length
scales. The
string mass per unit length obtained from COBE normalization is  
practically the same as in $\Omega_\Lambda=0$ case.

\begin{figure}[tbp]
\centerline{\epsfxsize = 0.86\hsize \epsfbox{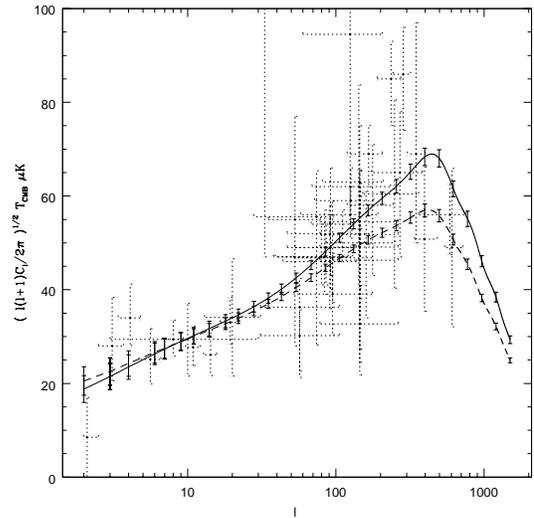}}
\vskip 0.5 truecm
\caption{
The total angular power spectrum for wiggly (solid line) and smooth 
(dashed line) strings when $\Omega_{baryons}=.05$, $\Omega_{CDM}=0.25$ and 
$\Omega_{\Lambda}=0.7$ and using small values for the string velocities:
$v=0.12$ in the radiation era and $0.1$ in the matter era.
}
\label{fig-c75}
\end{figure}

\begin{figure}[tbp]
\centerline{\epsfxsize = 0.86\hsize \epsfbox{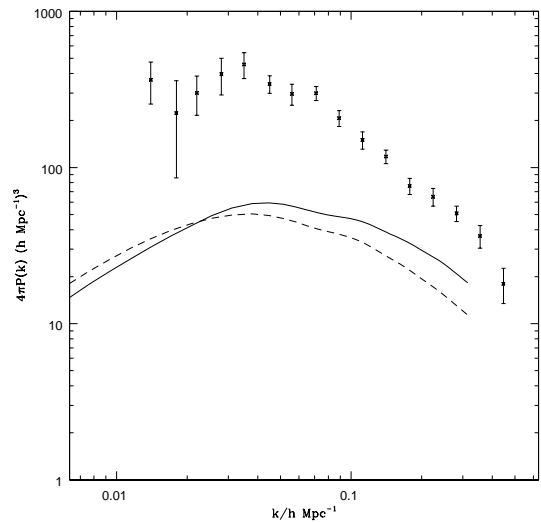}}
\vskip 0.5 truecm
\caption{
The matter power spectrum for
the same choice of parameters as in 
Fig. 
\protect \ref{fig-c75}.
}
\label{fig-p75}
\end{figure}

There is evidence from simulations \cite{shelal} that the large 
scale string velocities are much smaller than those plotted in Fig. \ref{fig-v}.
Also, wiggly strings must be heavier and slower than smooth
strings because of the different equation of state.  We have modified 
the parameters $c_r$, $c_m$, $k_r$ and $k_m$ of the one-scale
model in such a way that the rms velocity is $0.12$ in the radiation
era and $0.1$ in the matter era.
In Fig. \ref{fig-c75} and Fig. \ref{fig-p75} we plot the results of using
smaller velocities. We see a significant increase in the 
magnitude of the matter power spectrum. Including the wiggliness results
in a slight shift of power to smaller scales and an overall increase in 
magnitude. The scale-dependent bias, required to fit the data varies 
between $b\approx 1.6$ on smaller length scales and $b\approx 2.4$ on 
larger scales. COBE normalization gives 
$G\mu_0\approx 2.3\times 10^{-6}$ for smooth strings and 
$G\mu_0\approx 1.9\times 10^{-6}$ for wiggly strings.

\begin{figure}[tbp]
\centerline{\epsfxsize = 0.9\hsize \epsfbox{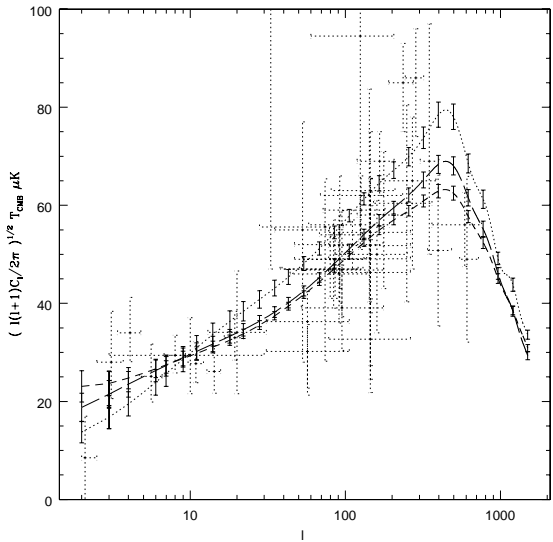}}
\vskip 0.5 truecm
\caption{
The total angular power spectrum for the network of wiggly strings
with $L=0.1$ (dotted line), $L=0.5$ (long dash line) and 
$L=0.92$ (short dash line). 
All other parameters are the same as in Fig.  \protect \ref{fig-c75}.
}
\label{fig-cL}
\end{figure}

\begin{figure}[tbp]
\centerline{\epsfxsize = 0.9\hsize \epsfbox{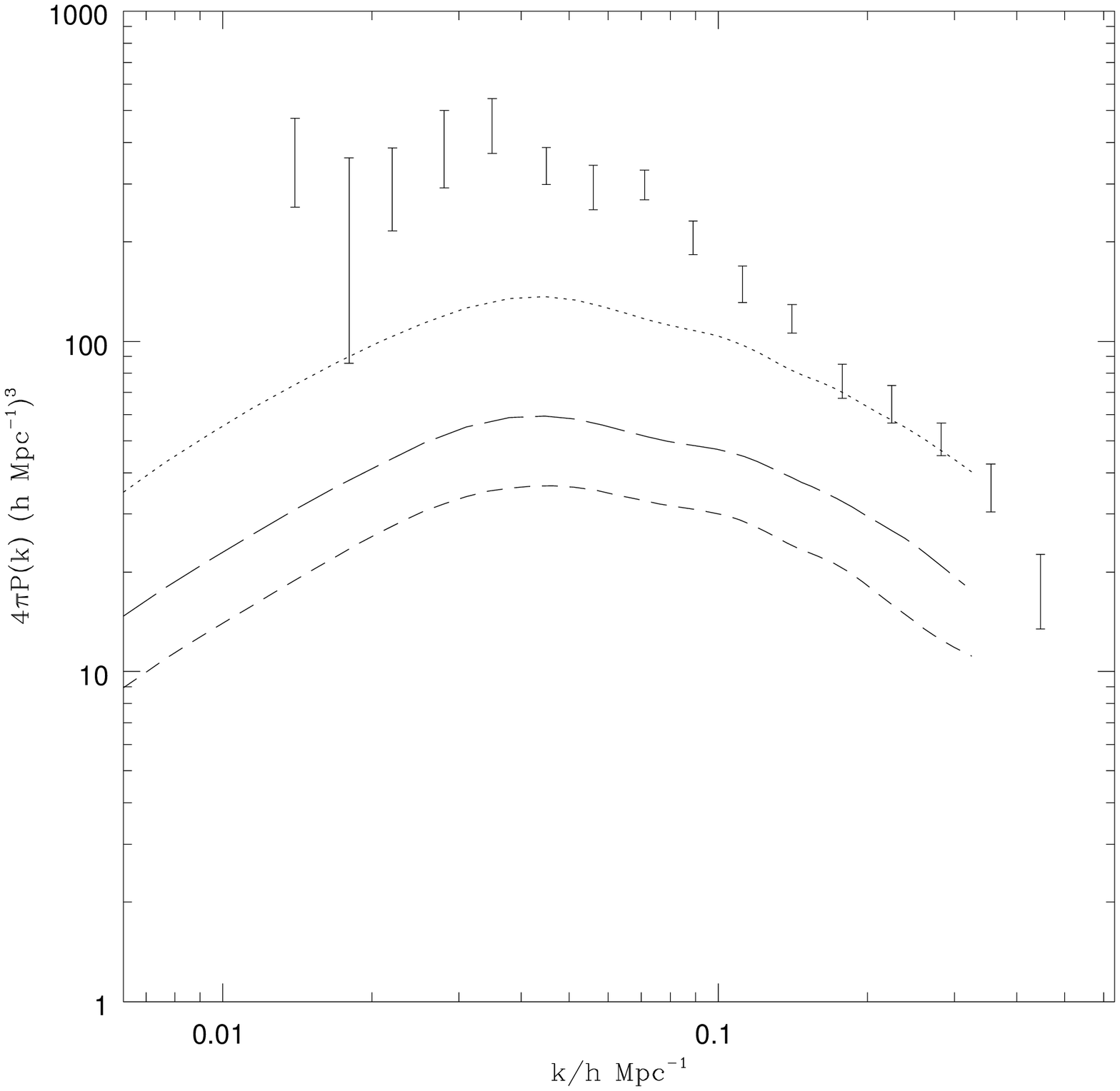}}
\vskip 0.5 truecm
\caption{
The matter power spectrum for
the same choice of parameters as in 
Fig. 
\protect \ref{fig-cL}.
}
\label{fig-pL}
\end{figure}

\begin{figure}[tbp]
\centerline{\epsfxsize = 0.9\hsize \epsfbox{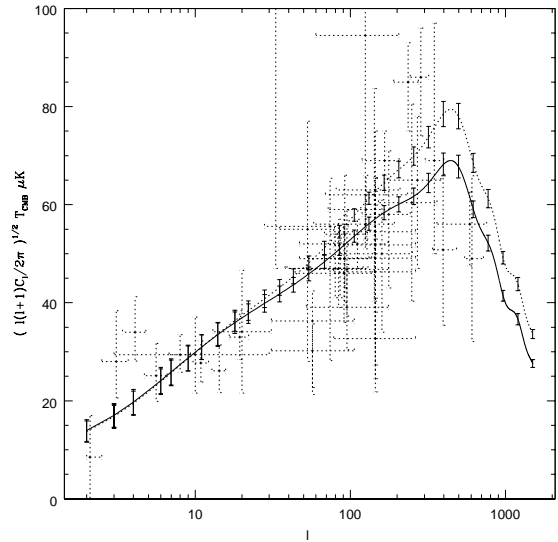}}
\vskip 0.5 truecm
\caption{
The total angular power spectrum for the network of wiggly strings
with $L=0.1$. The solid line corresponds to the modified ${\it l}(\tau)$. 
The dotted line is the same as in Fig. \protect \ref{fig-pL}. Other parameters
are the same as in Fig.\protect \ref{fig-c75}.
}
\label{fig-c}
\end{figure}

\begin{figure}[tbp]
\centerline{\epsfxsize = 0.9\hsize \epsfbox{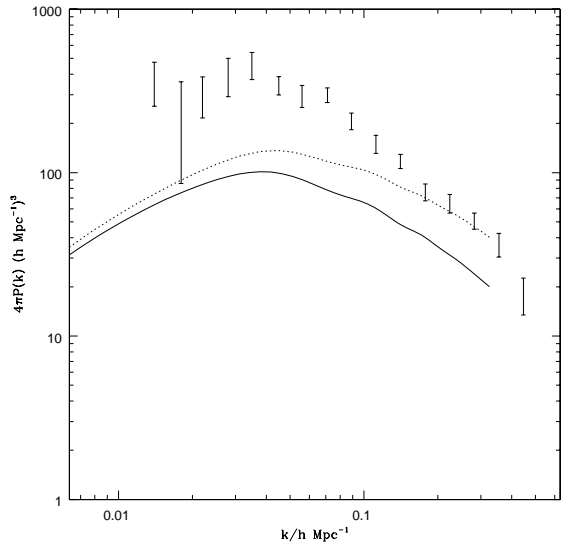}}
\vskip 0.5 truecm
\caption{
The matter power spectrum for
the same choice of parameters as in 
Fig. 
\protect \ref{fig-c}.
}
\label{fig-p}
\end{figure}

In addition, we have tested the dependence on two other parameters of the
model that may have physical significance. These are the string decay 
rate $L$ and the correlation length of the network 
${\it l}(\tau)$ . The parameter $L$ appears in the function $T^{off}$,
eq.(\ref{toff}), and controls how early and how fast the string segments 
decay. Smaller values of $L$ correspond to an earlier and slower decay. 
We have used $L=0.5$ for most of our calculations 
(Figures \ref{fig-c00}-\ref{fig-p75},\ref{fig-c},\ref{fig-p}).
In Fig. \ref{fig-cL} and Fig. \ref{fig-pL} we compare the results of 
using $L=0.1$, $0.5$ and $0.92$. Lower values of $L$ tend
to raise the matter power spectrum. When string segments start to decay 
earlier the effective string density at all times (and especially at later 
times) will decrease and a larger COBE normalization factor is required to 
fit the angular power spectrum on large scales. This, of course, encreases
the value of $G\mu_0$. We obtain $G\mu_0\approx 4\times 10^{-6}$ when using
$L=0.1$.

The evolution of the correlation length ${\it l}(\tau)$ 
(eq. (\ref{evol1}))
is completely determined by the parameters of the one-scale model
$c_r$, $c_m$, $k_r$ and $k_m$.
To obtain the results shown in Figs.
\ref{fig-c00}-\ref{fig-pL}
we have chosen these parameters so that the evolution of ${\it l}(\tau)$
is not altered and is the same as in Fig. \ref{fig-l}.
In Fig. \ref{fig-c} and Fig. \ref{fig-p} we plot the results of the 
simulation in which the cosmological parameters and
the string velocity were taken to be those in
Fig. \ref{fig-c75} but the parameters of
the one-scale model were set to give 
${\it l}(\tau)/\tau$ equal to $0.24$ in the radiation- and $0.3$ 
in the matter-dominated eras. This choice of parameters
is largely arbitrary, however, it shows that there is a freedom in
the model that may be used to change the shape of the matter
power spectrum.

\section{Conclusions}
\label{conclude}

We have seen that including the effects of small-scale structure in the 
string stress-energy improves the agreement with the observational data. 
Using $\Omega_\Lambda=0.7$ and reducing the average string velocities 
gives us a scale-dependent
bias factor between $1.6$ and $2.4$. Allowing for a slower decay
rate of strings and larger correlation length of the string network
make the bias factor nearly scale-invariant with an average value of $1.9$.
The scale-invariant bias factor of $2$ was reported in \cite{aletal2} which 
was obtained by introducing an additional time-dependence of $\mu$ designed 
to reproduce the shape of the matter power spectrum. The present work has the 
advantage of not altering the physics of the model.

The one-scale model, eqns. (\ref{evol1}) and (\ref{evol2}), is a very coarse 
approximation of the evolution of wiggly strings. The small-scale structure
is included only through the value of the effective curvature $\tilde{k}$,
the rest of the equations being exactly the same as for the smooth strings.
We expect that a more precise analytical model would improve the
agreement with the data. Also, work on improving the
model of the string network is in progress as reported in \cite{hind98}
and this should help to eliminate some of the assumptions we have had
to make in the present analysis. In addition we are presently investigating
ways to relax some of the assumptions of the model described in the
paper that will help shed light on the robustness of the predictions.


\

{\it Acknowledgements:} We would like to thank U. Seljak and M. Zaldarriaga
for making CMBFAST available to us and the National Energy Research Scientific 
Computing Center for the use of their J90 cluster. We are grateful to 
Martin White for discussion and Julian Borrill for help with supercomputing.
We thank Carlos Martins for pointing out the necessity of explaining 
our choice of the value for $g$.
 
TV was supported by a grant from the Department of Energy.

\end{document}